\documentclass[a4paper,12pt]{article}

\usepackage{a4,amsfonts,amsmath,amssymb}
\usepackage{cite}
\usepackage{url}
\RequirePackage{color}
\RequirePackage[colorlinks=true
,urlcolor=blue
,anchorcolor=blue
,citecolor=blue
,filecolor=blue
,linkcolor=blue
,menucolor=blue
,linktocpage=true
,pdfproducer=medialab
]{hyperref}
\usepackage{hyperref}
\newcommand{\be}{\begin{equation}}
\newcommand{\ee}{\end{equation}}
\newcommand{\nn}{\nonumber}

\newcommand{\D}{\mathcal{D}}

\newcommand{\LL}{\mathcal{L}}

\newcommand{\N}{\mathcal{N}}

\newcommand{\Sm}{\mathcal{S}}
\newcommand{\V}{\mathcal{V}}

\newcommand{\at}{\tilde{a}}

\newcommand{\et}{\tilde{e}}

\newcommand{\It}{\tilde{I}}
\newcommand{\Jt}{\tilde{J}}
\newcommand{\Kt}{\tilde{K}}

\newcommand{\xt}{\tilde{x}}
\newcommand{\yt}{\tilde{y}}

\newcommand{\dd}{\partial}

\newcommand{\al}{\alpha}
\newcommand{\bet}{\beta}
\newcommand{\ga}{\gamma}         \newcommand{\Ga}{\Gamma}
\newcommand{\de}{\delta}        
\newcommand{\ep}{\epsilon}

\newcommand{\la}{\lambda}       \newcommand{\La}{\Lambda}

\newcommand{\si}{\sigma}

\newcommand{\om}{\omega}

\title{
{\baselineskip -.2in
\vbox{\small\hskip 4in \hbox{IITM/PH/TH/2013/2}}
\vbox{\small\hskip 4in \hbox{IMSc/2013/06/05 \qquad  }}
} 
\vskip .4in
Stability of Bianchi attractors in Gauged Supergravity \\}
\date{}
\author{Karthik Inbasekar$^a$ and Prasanta K. Tripathy$^b$\\ \\
	\it $^a$Institute of Mathematical Sciences, \\
	\it CIT Campus, Taramani,\\
	\it Chennai 600113, India.\\
	\\ \\
	\it $^b$Department of Physics,\\
	\it Indian Institute of Technology Madras,\\
	\it Chennai 600 036, India.\\ \\
	email: $^a$ikarthik@imsc.res.in ; $^b$prasanta@physics.iitm.ac.in \\ \\}

\begin{document}

\maketitle

\begin{abstract}
In this paper, we analyse the stability of extremal black brane horizons with homogeneous symmetry in 
the spatial directions in five dimensional gauged supergravity, under the fluctuations of the scalar fields 
about their attractor values.  We examine the scalar fluctuation equations at the linearised level and 
demand that the fluctuations vanish as one approaches the horizon. Imposing certain restrictions on the 
Killing vectors used for gauging we find that the necessary conditions for stability are satisfied only by a 
subclass of the Bianchi metrics whose symmetry group 
factorises into a two dimensional Lifshitz symmetry and any homogeneous symmetry group given by 
the Bianchi classification. We apply these results to a simple example of a gauged supergravity model 
with one vector multiplet to find the stable attractors.
\end{abstract}

\newpage

\section{Introduction}

The study of extremal black branes in anti-deSitter space has become a topic of recent interest 
because of their appearance as candidates for gravity duals of field theories describing condensed
matter systems \cite{Kachru:2008yh}. Several examples of such solutions in the context of gravity
coupled to a dilaton as well as to various scalars and gauge fields have been studied
\cite{Goldstein:2010aw,Goldstein:2009cv}. 

In addition to their relevance as gravity duals to condensed matter field theories, such black
brane geometries also appear in natural extensions of the attractor mechanism to gauged
supergravities \cite{Ceresole:2001wi,Cacciatori:2009iz,Dall'Agata:2010gj,Hristov:2010ri,
Barisch:2011ui,BarischDick:2012gj,Hristov:2012nu}, there by providing a larger class of
configurations for studying the thermodynamics of black holes. A classification of such
configurations which are homogeneous but not isotropic has also been carried out in
\cite{Iizuka:2012iv,Iizuka:2012pn}. These attractor geometries, known as the Bianchi attractors,
have a richer structure. They are characterised by constant anholonomy coefficients and are regular.
In a number of cases belonging to Bianchi type I, they can be obtained either from a gauged
supergravity theory or from string theory by a suitable choice of the internal manifold for
compactification\cite{Balasubramanian:2010uk,Cassani:2011sv,Halmagyi:2011xh}. A prescription has
been given to obtain such generalised attractors from gauged supergravity \cite{Kachru:2011ps}. In
this framework one sets all the fields and the curvature components to constants in the tangent
space. 
Following this prescription, in a previous work \cite{Inbasekar:2012sh}, we embedded some of the
Bianchi attractors \cite{Iizuka:2012iv} in five dimensional gauged supergravity. We considered a
simple gauged supergravity model and constructed Lifshitz, Bianchi II and Bianchi VI solutions. 

One of the important issues being investigated currently is the stability of such Lorentz
violating geometries \cite{Copsey:2010ya,Keeler:2012mb,Harrison:2012vy,Bao:2012yt,Andrade:2012xy,
Andrade:2013wsa}. Instabilities due to scalar field fluctuations were found to exist in a class of
charged black brane geometries \cite{Donos:2011qt,Cremonini:2012ir}. Presence of such instabilities
in these solutions plays a crucial role because they indicate that the geometry might get corrected
in the deep infrared \cite{Harrison:2012vy}. Though the stability analysis has been carried out in a
number of examples, a common recipe (to figure out whether certain geometry has any instability) is
still lacking. 

Though the attractor mechanism \footnote{See
\cite{Ferrara:1995ih,Strominger:1996kf,Ferrara:1996dd,Ferrara:1997tw} for some early works and the
reviews \cite{Ferrara:2008hwa,Bellucci:2007ds} for a detailed analysis of the attractor
mechanism.} has been studied quite extensively in the context of extremal black holes 
in Minkowski space with near horizon geometry $AdS_2 \times S^2$, the study of generalised
attractors has not yet been explored thoroughly for the new class of Lorentz violating geometries
arising as gravity duals of condensed matter systems. Especially, it is not at all obvious which
among these entire class of new attractor geometries are stable and can survive in the deep
infrared. Since a number of such geometries can be embedded in gauged supergravity, where the 
scalar couplings and potential term are determined by symmetry, it is natural to ask whether these gauged 
supergravity attractors are stable.

In this paper, we analyse the stability of electrically charged Bianchi attractors in gauged
supergravity. For attractors which asymptote to Minkowski space the conditions for stability is well
understood \cite{Goldstein:2005hq}. In such cases the attractor values of the scalar 
fields must correspond to an absolute minimum of the black hole potential. In the present paper we
derive the analogous condition for the generalised attractors. We consider scalar fluctuations about
the attractor value and analyse the scalar field equations in the background of the Bianchi
geometries. We take the fluctuations to depend on time and on the radial direction as we are mainly
interested in determining the radial behaviour. We study the stress energy tensor in gauged
supergravity and expand it in first order of scalar fluctuations. We find that the stress energy
tensor in gauged supergravity depends on the scalar fluctuations even at first order perturbation
due to non-trivial interaction terms in the theory. If there is a large backreaction due to scalar
fluctuations, the geometry would significantly differ from the attractor geometry indicating an
instability. Therefore, stable attractor geometries are those where the scalar fluctuations die out
as one approaches the horizon.

We then study the scalar field equations with the fluctuations at first order, determine the
general solution and the conditions under which these fluctuations can exist. These conditions are
such that the generalised attractor geometries must exist at critical points which are maxima of the
attractor potential. We then derive conditions for stability of the Bianchi attractors in gauged
supergravity by studying the near horizon behaviour of the scalar fluctuations and demanding
regularity. In particular, we find that this severely restricts the general form of
these metrics.{\footnote{In deriving this result, we make certain technical assumption on the killing 
vectors used in gauging, as well as on the nature of the critical points giving rise to the attractor 
geometry which will be discussed in due course.}} We find that metrics which factorise as
\be\label{met}
ds^2=L^2 \biggl[-\hat{r}^{2u_0} d\hat{t}^2+ \frac{d\hat{r}^2}{\hat{r}^2}+\eta_{ij}\om^i\otimes
\om^j\biggr],
\ee
are stable under scalar fluctuations about the attractor value. The parameter $u_0$ must be 
positive in order to have a regular horizon, $i=1,2,3$ corresponds to the $\hat{x},\hat{y},\hat{z}$
directions, $\eta_{ij}$ is a constant matrix independent of co-ordinates and $\om^i$ are the one
forms invariant under the homogeneous symmetries.\footnote{For more details on the homogeneity and
invariant one forms, see \cite{Iizuka:2012iv,ryan1975homogeneous,1975classical}.} We call this
special sub class of metrics as $Lif_{u_0}(2) \times M $.\footnote{The notation $Lif_{u_0}(d)$ has
been used in the literature to denote the d dimensional Lifshitz metric with exponent $u_0$. The
three dimensional submanifold $M$ is constructed from one forms invariant under the homogeneous
groups given by the Bianchi classification, there are 9 such submanifolds each denoted as $M_I,
M_{II},\ldots M_{IX}$.} Note that the symmetry group of \eqref{met} factorises into a
direct product of a 1+1 dimensional non-relativistic conformal group times a homogeneous group
of symmetries in three dimensions.
The metric has the scaling symmetry, 
\be 
\hat{r}\rightarrow \al \hat{r}, \quad \hat{t}\rightarrow \frac{\hat{t}}{\al^{u_0}},
\ee
and homogeneous symmetries generated by the Bianchi groups along the $\hat{x},\hat{y},\hat{z}$
directions. When $u_0=1$, we get an $AdS_2$ factor and the symmetry is enhanced to $SO(2,1) \times
M$. This factorisation is reminiscent of extremal black holes in four dimensions where the near
horizon geometries factorise as $AdS_2\times S^2$.

The organisation of the paper is as follows. In \S\ref{gaugedsugraandgeneralisedattractors} we
give a brief background on gauged supergravity and generalised attractors. Subsequently, in
\S\ref{gaugefieldfluctuation} we describe the field equations satisfied by the gauge field
fluctuations. Following which, we describe the scalar fluctuations about the attractor values
followed by an expansion of the stress energy tensor with the fluctuations in
\S\ref{stressenergyanalysis}. We then derive the general solutions for the scalar fluctuations and
describe the conditions under which these fluctuations exist in \S\ref{scalarfieldanalysis}.
Following this we study the near horizon behaviour of the fluctuations and derive stability
conditions for the Bianchi attractors in \S\ref{stablebianchiattractors}. In appendices
\S\ref{tangentspaceandanholonomy} and \S\ref{gauged_sugra_model} we give some background material on
a simple gauged supergravity model. In \S\ref{gauged_sugra_solutions} we have summarised some
examples of Bianchi attractors in gauged supergravity from our earlier work. Finally, in
\S\ref{ads2R3ingaugedsugra} we have provided some new examples of $Lif_{u_0}(2) \times M$ class of
metrics in a $U(1)_R$ gauged supergravity.

\section{Gauged supergravity and generalised attractors}\label{gaugedsugraandgeneralisedattractors}

In this section, we briefly summarise some necessary background in gauged supergravity
and generalised attractors. We switch off the tensor as well as the hyper multiplets and turn on
only the vector multiplets. The bosonic part of the $\N=2, d=5$ gauged supergravity is then given by
by \cite{Ceresole:2000jd}:
\begin{align}\label{lagrangian}
\hat{e}^{-1} \LL^{\N=2}_{Bosonic}= & -\frac{1}{2}R -\frac{1}{4} a_{I J}F_{\mu\nu}^I F^{J
\mu\nu}-\frac{1}{2}g_{x y}\D_\mu
\phi^x\D^\mu\phi^y\nn\\
&-\V(\phi)+\frac{\hat{e}^{-1}}{6\sqrt{6}}C_{IJK}\epsilon^{\mu\nu\rho\sigma\tau}
F^I_{\mu\nu}F^J_{\rho\sigma}A^K_\tau \ .
\end{align}
Here $\hat{e}= \sqrt{-det g_{\mu\nu}}$ and $a_{IJ}$ is the ambient metric used to raise and lower the
vector indices and $g_{xy}$ is the metric on the moduli space. The moduli space spanned by the
real scalar fields $\phi^x$ is a very special manifold. It is completely specified in terms of 
constant symmetric tensors $C_{IJK}$ by the hypersurface:
\be
C_{IJK} h^I h^J h^K=1, \quad h^I\equiv h^I(\phi).
\ee
The group of isometries of this manifold is 
denoted by $G$. The gauging is specified in terms of a subgroup $K\subset G$ generated by the 
Killing vectors $K_I^x$. The covariant derivatives on the scalars are given in terms of the Killing vectors
$K_I^x$ as:
\be
\D_\mu\phi^{x} \equiv \dd_\mu\phi^{x} + g A^I_\mu K_I^{x}(\phi) \ .
\ee
%
The potential term is given by,
\be\label{potential}
\V(\phi)=-g_R^2 [ 2P_{ij}P^{ij}-P^{\at}_{ij}P^{\at ij} ],
\ee
where $g_R$ is the gauge coupling constant associated with gauging the R symmetry group and
\be
 P_{ij} \equiv h^I P_{Iij}, \quad P^{\at}_{ij} \equiv h^{\at I}P_{Iij}.
\ee
Here $i,j=1,2$, $\at=0,1,\ldots n_V$ and for the case of $U(1)_R\subset SU(2)_R$ gauging,
$P_{Iij}=V_I \de_{ij}$. 

In the following, we give  various field equations of this theory that are necessary for
the stability analysis. All the equations are written in position space for the purpose of this
paper. It was shown in \cite{Inbasekar:2012sh} that all the field equations become algebraic at the
attractor points defined by,
\be
\phi^{x}={\rm const} \ ; A^I_a={\rm const} \ ; c_{bc}^{\ \ a}={\rm const},
\ee
where $c_{bc}^{\ \ a}$ are called as the anholonomy coefficients (see
\S\ref{tangentspaceandanholonomy} for a detailed discussion).

The gauge field equation is:
\be\label{gaugefieldequation}
\dd_\mu(\hat{e}a_{IJ} F^{I\mu\nu})=\hat{e}(g^2K_{IJ}A^{\nu J}+g K_{Iy} \dd^{\nu}\phi^y),
\ee
where we have defined $K_{IJ}=K_I^x K_J^y g_{x y}$. Note that we have ignored the Chern-Simons
term, since it vanishes for the Bianchi attractors \cite{Inbasekar:2012sh}. At the attractor
points, the equation simplifies to,
\be\label{gaugefieldattractoreq}
a_{IJ}|_{\phi_c}\nabla_\mu F^{I \mu\nu}= g^2 K_{IJ}|_{\phi_c}A^{\nu J}
\ee

The equation of motion for the scalar fields $\phi^x$ is given by:
\begin{align}\label{scalareq}
\hat{e}^{-1}\dd_\mu \big[\hat{e} \ g_{z y} \D^\mu \phi^{y} \big] &-\frac{1}{2} \frac{\dd g_{x
y}}{\dd\phi^{z}}\D_\mu\phi^{x}\D^\mu\phi^{y}-g A_\mu^I g_{x y}\frac{\dd
K^{x}_I}{\dd\phi^{z}}\D^\mu\phi^{y}\nn \\
&-\frac{1}{4}\frac{\dd a_{I J}}{\dd\phi^{z}}F^{I}_{\mu\nu} F^{J
\mu\nu}-\frac{\dd\V(\phi)}{\dd\phi^{z}}=0 \ , 
\end{align}
where as the Einstein's equation is:
\be
R_{\mu\nu}-\frac{1}{2}R g_{\mu\nu}=T_{\mu\nu}.
\ee
The stress energy tensor $T_{\mu\nu}$ has the expression: 
\begin{align}\label{stressenergy}
T_{\mu\nu}= & g_{\mu\nu} \biggl[\frac{1}{4} a_{I J}F_{\mu\nu}^I F^{J
\mu\nu}+\frac{1}{2}g_{x y}\D_\mu \phi^x\D^\mu\phi^y+\V(\phi)\biggr]\nn\\
&- \biggl[a_{I J} F^I_{\mu\la}F^{J \ \la}_\nu+ g_{x y}\D_\mu\phi^x \D_\nu\phi^y\biggr].
\end{align}
At the attractor points the scalar field equation reduces to minimisation of the attractor potential,
\be\label{attractorequation}
\frac{\dd\V_{attr}(\phi,A)}{\dd\phi}=0,
\ee
which has the expression:
\be\label{attractorpotential}
\V_{attr}(\phi,A)=\V(\phi)+\frac{1}{2}g^2 K_{IJ} A_\mu^I A^{\mu J}+\frac{1}{4}a_{IJ}
F^{I \mu \nu}F^J_{\mu\nu}.
\ee

The critical points of the attractor potential are denoted as
$\phi_c$, and solving \eqref{attractorequation} relates $\phi_c$ in terms of the charges. In
\cite{Inbasekar:2012sh} we considered a simple gauged supergravity model with one vector multiplet
and constructed examples of Bianchi attractors. Some details of the gauged supergravity model and 
the Bianchi attractor solutions are provided in appendices \S\ref{gauged_sugra_model} and
\S\ref{gauged_sugra_solutions} respectively.
\section{Scalar Perturbation about attractor values}\label{scalarperturbation}
In this section, we consider the fluctuations of the scalar fields about their attractor value. We
take the fluctuation to be of the form,
\be\label{scalarfluctuation}
\phi_c+ \ep\de\phi(r,t),
\ee
where $t$ denotes the time, $r$ is the radial direction, $\phi_c$ are the attractor values of
the scalars and $\de\phi$ is the perturbation with $\ep<1$. We have taken the
fluctuation to not depend on the $(x,y,z)$ directions to respect the Bianchi type symmetries along
these directions. Besides, we are primarily interested in the radial behaviour of the fluctuation as
one approaches the horizon. We also assume that the black brane metric can be expanded about the
near horizon geometries as follows,
\be
\tilde{g}_{\mu\nu} \sim g_{\mu\nu}(r-r_h)+ \epsilon \ \ga_{\mu\nu}(r-r_h)+ O(\epsilon^2)+\ldots ,
\ee
where $g_{\mu\nu}$ is the near horizon metric given by the Bianchi type geometries.
The higher order terms like $\ga_{\mu \nu}$ are due to the back reaction of the scalar field
fluctuations on the attractor geometry. 
\subsection{Gauge field fluctuations}\label{gaugefieldfluctuation}
We also consider gauge field fluctuations together with scalar and metric fluctuations. In this
subsection, we derive the gauge field fluctuation equation in terms of scalar fluctuations. Let us
consider fluctuations of the form
\be
A_\mu^I+\ep\de A_\mu^I,
\ee
where $A_\mu^I$ are the attractor values of the gauge field for any given Bianchi metric. The field
strength expands as,
\be
F_{\mu\nu}^I+ \ep F_{\mu\nu}^{I \de}, \quad F_{\mu\nu}^{I \de}= \dd_\mu\de A_\nu^I-\dd_\nu\de
A_\mu^I
\ee
We now expand the scalars and gauge fields about the attractor value in
\eqref{gaugefieldequation} to get,
\begin{align}
 a_{IJ}|_{\phi_c}\nabla_\mu F^{I\mu\nu}_\de-g^2 K_{IJ}|_{\phi_c} \de A^{\nu J}=&-\biggl( \frac{\dd
a_{IJ}}{\dd\phi^z}\bigg|_{\phi_c} \nabla_\mu(F^{I \mu\nu}\de\phi^z)-g^2
\frac{\dd K_{IJ}}{\dd\phi^z}\bigg|_{\phi_c} \de\phi^z A^{\nu J}\biggr) \nn \\
&+g K_{Iy}|_{\phi_c} \dd^\nu \de\phi^y,
\end{align}
where we have used \eqref{gaugefieldattractoreq} for simplification. Note that we did not have to
consider metric fluctuations in the above equation, since it would lead to second order terms. It
can be seen that regular behaviour of the gauge field fluctuations depend on regularity of the
scalars and their derivatives near the horizon. In other words, the gauge fluctuation is not 
independent of the scalar fluctuation but varies as $\delta A^{\mu I} \sim \delta\phi^x$. Subsequently
in \S3.3 we will argue that, for stable attractors the scalar field fluctuation $\delta\phi^x$ must vanish 
at the attractor point. The above analysis shows that the attractor which is stable against scalar 
field fluctuation is also stable against gauge field fluctuation. 

\subsection{Backreaction at first order}\label{stressenergyanalysis}
We now expand the stress energy tensor \eqref{stressenergy} up to first order in $\ep$ under the
scalar perturbations \eqref{scalarfluctuation}, 
and simplify, to get:
\begin{align}
T_{\mu\nu}(\phi_c+\de\phi)= & T_{\mu\nu}^{attr}|_{\phi_c} + g
K_{yI}|_{\phi_c}\Big(A^{\la I}\dd_\la(\de\phi^y) g_{\mu\nu}- A_\mu^I\dd_\nu (\de\phi^y)
 - A_\nu^I\dd_\mu(\de\phi^y)\Big)\nn\\
& +\biggl[\frac{\dd a_{IJ}}{\dd\phi^z}\bigg|_{\phi_c}\biggl(\frac{1}{4} g_{\mu\nu}
F^{\la\si I} F^J_{\la\si}-F^I_{\mu\la}F^{J \ \la}_\nu\biggr)\nn\\
& +g^2\frac{\dd K_{IJ}}{\dd\phi^z}\bigg|_{\phi_c}\biggl(\frac{1}{2} g_{\mu\nu} A_\la^I A^{\la
J}-A_\mu^I A_\nu^J\biggr)+\frac{\dd\V}{\dd\phi^z}\bigg|_{\phi_c}\biggr]\de\phi^z \ , 
\end{align}
where we have introduced
\be
T_{\mu\nu}^{attr}|_{\phi_c}=\V_{attr}(\phi_c)g_{\mu\nu} - \bigg[a_{IJ}|_{\phi_c} F_{\mu \la}^I
F_{\nu}^{\ \la J}+ g^2 K_{IJ}|_{\phi_c} A_\mu^I A_\nu^J \bigg].
\ee
The attractor equations \eqref{attractorequation} can be used for further simplification to get,
\begin{align}\label{stressenergyfirstorder}
T_{\mu\nu}(\phi_c+\de\phi)= & T_{\mu\nu}^{attr}|_{\phi_c} + g
K_{yI}|_{\phi_c}\Big(A^{\la I}\dd_\la(\de\phi^y) g_{\mu\nu}- A_\mu^I\dd_\nu(\de\phi^y)
- A_\nu^I \dd_\mu(\de\phi^y)\Big)\nn\\
& -\biggl[\frac{\dd a_{IJ}}{\dd\phi^z}\bigg|_{\phi_c}F^I_{\mu\la}F^{J \
\la}_\nu+g^2\frac{\dd K_{IJ}}{\dd\phi^z}\bigg|_{\phi_c}A_\mu^I A_\nu^J\biggr]\de\phi^z.
\end{align}
It is already clear that for general perturbations of the scalar field, there is backreaction at
first order even after using the attractor equations. In particular this requires the fluctuations
and their derivatives to be well behaved as one approaches the horizon. Any divergent fluctuation
would cause infinite backreaction and deviation from the attractor geometry indicating an
instability. Taking the trace of \eqref{stressenergyfirstorder} we get,
\begin{align}
T_\mu^{\ \mu}(\phi_c+\de\phi)=& T_{\mu}^{attr \mu}|_{\phi_c}+ (d-2) g
K_{yI}|_{\phi_c}A^{\la I}\dd_\la(\de\phi^y)\nn\\
&-\biggl[\frac{\dd
a_{IJ}}{\dd\phi^z}\bigg|_{\phi_c}F^I_{\mu\nu}F^{J\mu\nu}+g^2\frac{\dd
K_{IJ}}{\dd\phi^z}\bigg|_{\phi_c}A_\mu^I A^{J\mu}\biggr]\de\phi^z,
\end{align}
where $d$ is the space time dimension. Once again we can use the attractor equations
\eqref{attractorequation} to simplify, and the Einstein equations take the form,
\begin{align}
R \frac{(2-d)}{2}=& T_{\mu}^{attr \mu}|_{\phi_c}+ (d-2) g
K_{yI}|_{\phi_c}A^{\la I}\dd_\la(\de\phi^y)\nn\\
&+\biggl[g^2\frac{\dd K_{IJ}}{\dd\phi^z}\bigg|_{\phi_c}A_\mu^I
A^{J\mu}+4\frac{\dd\V}{\dd\phi^z}\bigg|_{\phi_c}\biggr]\de\phi^z.
\end{align}
Suppose the critical points of the attractor potential are also simultaneous critical points
of the gauged supergravity scalar potential (as was the case with all the examples discussed in
\cite{Inbasekar:2012sh}), we see that the terms relevant for the backreaction are proportional to
$g$: 
\be\label{stressenergytracegeneralgauging}
R \frac{(2-d)}{2}= T_{\mu}^{attr \mu}|_{\phi_c}+ (d-2) g
K_{yI}|_{\phi_c}A^{\la I}\dd_\la(\de\phi^y)+g^2\frac{\dd
K_{IJ}}{\dd\phi^z}\bigg|_{\phi_c}A_\mu^I
A^{J\mu}\de\phi^z,
\ee
Thus, for gauging of $R$ symmetry, $g=0$ and hence the backreaction is absent:
\be\label{stressenergytraceRsymmetrygauging}
R \frac{(2-d)}{2}= T_{\mu}^{attr \mu}|_{\phi_c}.
\ee
(See \S\ref{ads2R3ingaugedsugra} for some examples of generalised attractor in gauged 
supergravity with just R symmetry gauging).  However, in gauged supergravity with a 
generic gauging of symmetries of the scalar manifold, the equation depends on the
first order fluctuations in the scalar fields. Thus, the generalised attractor geometries in gauged
supergravity with a generic gauging can get backreacted by fluctuations of scalar fields. It then
follows that the relevant boundary conditions to have stable attractors should be such that the
fluctuations and derivatives of fluctuations vanish as one approaches the horizon.

The main point of the above calculation was to indicate that there are first order scalar
fluctuation terms present in the expansion of the stress energy tensor. We now include the metric
fluctuation as well and write down the linearised Einstein equation: 
\begin{align}\label{linearisedeq}
 \nabla^\al \nabla_\al \bar{\ga}_{\mu\nu}+ 2 R_{(\mu \ \nu) \ }^{\ \ \al \
\ \beta}\bar{\ga}_{\beta\al} - 2 R_{(\mu}^{ \ \ \beta} \bar{\ga}_{\nu)\beta}+g_{\mu\nu}
(R_{\al\beta}\bar{\ga}^{\al\beta}+\frac{2}{2-D}R \bar{\ga})+ R\bar{\ga}_{\mu\nu} \nn\\ + 2
\dot{T}_{\mu\nu}^{attr}(g_{\al\beta}+\ep\ga_{\al\bet})|_{\ep=0}=0 \ . 
\end{align}
Our conventions are as in \cite{wald1984general}. We denote $\gamma_{\mu\nu}$ to be the first order 
perturbation about the metric, the dot indicates derivative w.r.t $\ep$,  
$\bar{\ga}_{\mu\nu}=\ga_{\mu\nu}-\frac{1}{2}\ga
g_{\mu\nu}$, $\bar{\ga}=\frac{2-D}{2}\ga$, and $\ga=g^{\mu\nu}\ga_{\mu\nu}$. All the covariant
derivatives, Riemann tensor, Ricci tensor and curvature terms are with respect to the unperturbed
metric $g_{\mu\nu}$. The terms in the stress energy tensor dependent on the metric fluctuations are
given by,
\begin{align}\label{stressenergymetricdependence}
\dot{T}_{\mu\nu}^{attr}(g_{\al\beta}+\ep\ga_{\al\bet})|_{\ep=0}=& \V_{attr}(\phi_c)
(\bar{\ga}_{\mu\nu}+\frac{2\bar{\ga}}{2-D}g_{\mu\nu})\nn\\
&-(\bar{\ga}_{\la\si}+\frac{\bar{\ga}}{2-D} g_ { \la\si})
(\frac{1}{2}T_{attr}^{\la\si}g_{\mu\nu}+a_{IJ}|_{\phi_c} F^{I \ \la}_{\mu}F^{J \ \si}_{\nu}).
\end{align}
To include the scalar fluctuations one just need to add \eqref{stressenergyfirstorder} to
\eqref{stressenergymetricdependence}. The condition of stability boils down to the existence of a 
one parameter family of solutions to \eqref{linearisedeq} (see chapter 8 of  \cite{wald1984general}).
The equation \eqref{linearisedeq} is a very complicated set of differential equations and exact 
analysis is possible only in the case of a flat background or for a very specialised set of
background
metric  \cite{wald1984general}. Proof of existence of a one parameter family of solutions in our attractor 
background is 
beyond the scope of the current analysis and we leave a detailed analysis of  this for future work.

\subsection{Scalar fluctuations}\label{scalarfieldanalysis}

In this section, we will analyse the scalar fluctuations in detail using the equation of motion for 
the scalar fields. The field equation \eqref{scalareq}  can be rewritten as:
\be
\hat{e}^{-1}\dd_\mu \big[\hat{e} \ g_{z y} \D^\mu \phi^{y} \big] -\frac{1}{2} \frac{\dd g_{x
y}}{\dd\phi^{z}}\nabla_\mu\phi^{x}\nabla^\mu\phi^{y}- g \frac{\dd K_{I y}}{\dd\phi^z} A_\mu^I
\nabla^\mu\phi^y-\frac{\dd\V_{attr}} {\dd\phi^z}=0.
\ee
We will now expand the scalar fields about their attractor values  and keep terms of $O(\ep)$ to get:
\begin{align}
 g_{zy}|_{\phi_c} \nabla_\mu \nabla^\mu
\de\phi^y-\frac{\dd^2\V_{attr}}{\dd\phi^z\dd\phi^y}\bigg|_{\phi_c}\de\phi^y &+ g \biggl[\frac{\dd
K_{I z}}{\dd\phi^y}-\frac{\dd K_{I y}}{\dd\phi^z}\biggr]\bigg|_{\phi_c}A^{\mu
I}\nabla_\mu \de\phi^y\nn\\
&+g\biggl[ K_{I z}|_{\phi_c}+ \frac{\dd K_{I
z}}{\dd\phi^y}\bigg|_{\phi_c}\de\phi^y\biggr] \nabla_\mu A^{\mu I}=0.
\end{align}
Here the covariant derivative $\nabla_\mu$ is taken with respect to the zeroth order metrics which
represent  the
near horizon Bianchi geometries. Note that the higher order metric terms which are undetermined 
are not required at $O(\ep)$. We choose the gauge condition $\nabla_\mu A^{\mu I}=0$ to eliminate
the last term. Finally we get,
\be\label{fluctuationeq1}
\nabla_\mu \nabla^\mu
\de\phi^x- g^{zx}\frac{\dd^2\V_{attr}}{\dd\phi^z\dd\phi^y}\bigg|_{\phi_c}\de\phi^y + 2 g
\ (g^{zx} \tilde{\nabla}_y K_{I z})|_{\phi_c}A^{\mu I}\nabla_\mu \de\phi^y=0,
\ee
where $\tilde{\nabla}$ is the covariant derivative with respect to the metric on the scalar manifold
$g_{xy}$. The Laplacian operator can be written as,
\be\label{laplacian1}
\nabla_\mu \nabla^\mu =g^{\hat{r}\hat{r}} \dd^2_{\hat{r}} + g^{\hat{t}\hat{t}}\dd^2_{\hat{t}} +
(g^{\hat{r}\hat{r}}\frac{\dd_{\hat{r}} \hat{e}}{\hat{e}}+\dd_{\hat{r}} g^{\hat{r}\hat{r}}
)\dd_{\hat{r}},
\ee
since the scalar fluctuations depend only on the radial and time co-ordinates. 

Before substituting
the details we would like to make some comments on the co-ordinate system used
for writing the Bianchi attractor geometries. In \cite{Iizuka:2012iv} the horizon for the Bianchi
metrics was located at $r=-\infty$,  where as in \cite{Inbasekar:2012sh} we have chosen the 
co-ordinate $\hat{r}= e^r$ such that the horizon lies at $\hat{r}=0$ instead (also see 
\S\eqref{gauged_sugra_solutions}). As can be seen from the general form of the Bianchi metrics, 
\be\label{bianchimetricgeneralform}
ds^2=L^2 \biggl[-\hat{r}^{2u_0} d\hat{t}^2+ \frac{d\hat{r}^2}{\hat{r}^2}+ \hat{r}^{2 (u_i+u_j)}
\eta_{ij}\om^i\otimes \om^j\biggr],
\ee
the constants $u_0, u_i$ must be positive in order to have a regular horizon. Here the index
 $i=1,2,3$ corresponds to the $\hat{x},\hat{y},\hat{z}$ directions, $\eta_{ij}$ is a constant matrix 
 independent of the co-ordinates and $\om^i$ are the one forms invariant under the 
 homogeneous symmetries \cite{Iizuka:2012iv}. 
 
 Thus one can see that the general form  of the determinant is
\be
\hat{e}=\sqrt{-det g_{\mu\nu}}\sim L^5 \hat{r}^m f(x,y,z),
\ee
where $m=-1+\sum_l c_l u_l$ , $u_l$ are the various exponents and $c_l$ is a positive number
with $c_0=1$ for all Bianchi attractors. For example, in the Bianchi II case (see \eqref{solutionII})
$m=-1+u_0+2(u_1+u_2)$. We can also see
that,
\be
g^{\hat{r}\hat{r}}=\frac{\hat{r}^2}{L^2} ,\quad g^{\hat{t}\hat{t}}=-\frac{1}{L^2 \hat{r}^{2u_0}},
\ee
for all Bianchi attractors. Using the above data, the Laplacian \eqref{laplacian1} can be expressed
as,
\be\label{laplacian2}
\nabla_\mu
\nabla^\mu=\frac{1}{L^2}\biggl[\hat{r}^2\dd^2_{\hat{r}}+(m+2)\hat{r}\dd_{\hat{r}}-\frac{1}{\hat{r}^{
2u_0}}\dd^2_{\hat{t}} \biggr].
\ee
Substituting \eqref{laplacian2} in \eqref{fluctuationeq1} and using the ansatz
\eqref{gaugefieldansatz}
for $A_\mu^I$ we get,
\be
\biggl[\hat{r}^2\dd^2_{\hat{r}}+(m+2)\hat{r}\dd_{\hat{r}}-\frac{1}{\hat{r}^{
2u_0}}\dd^2_{\hat{t}} \biggr]\de\phi^x-
M^x_y|_{\phi_c}\de\phi^y+N^x_y|_{\phi_c}\frac{1}{\hat{r}^{u_0}}\dd_{\hat{t}} \de\phi^y=0,
\ee
where,
\be\label{definitions}
M^x_y|_{\phi_c}=L^2 g^{zx}\frac{\dd^2\V_{attr}}{\dd\phi^z\dd\phi^y}\bigg|_{\phi_c}, \quad
N^x_y|_{\phi_c} =2 g L A^{I \bar{0}} (g^{zx} \tilde{\nabla}_y K_{I z})|_{\phi_c}.
\ee

The metric on the moduli space $g_{xy}$ is chosen to be positive definite and the nature of the critical point is given
by the sign of the double derivative of the attractor potential. We further assume that
$M^x_y|_{\phi_c}$ is diagonal so that, 
\be\label{doublederivativeofattractorpotential}
M^x_y|_{\phi_c}\de\phi^y=\la \de\phi^x.
\ee 
The term $N^x_y$ can be non zero in general, but vanishes trivially for the gauged supergravity
model where we found some examples  Bianchi attractors (see \S\eqref{gauged_sugra_model}). 
There is only one
Killing vector \eqref{killingvectorSO2} that generates the $SO(2)$ isometry on the scalar manifold,
and the critical point is such that $\phi^2_c=\phi^3_c=0$. Therefore one is left with just the
$\tilde{\nabla}_x K_{I x} $ component which vanishes due to the Killing vector equation on the
manifold.\footnote{Here, the single surviving component of the Killing vector is along the direction
of $\phi^1$ on the scalar manifold.}

Thus, the scalar fluctuation equation \eqref{fluctuationeq1} has the final form,
\be
\biggl[\hat{r}^2\dd^2_{\hat{r}}+(m+2)\hat{r}\dd_{\hat{r}}-\frac{1}{\hat{r}^{
2u_0}}\dd^2_{\hat{t}} - \la\biggr]\de\phi^x=0.
\ee

The above equation admits a simple solution when the fluctuations $\delta\phi^x$ are 
time independent. In this case, we have 
\be
\delta\phi^x = C_1 r^{\big(\sqrt{4\lambda + (1+m)^2}- (1+m)\big)/2} 
+ C_2  r^{\big(-\sqrt{4\lambda + (1+m)^2}- (1+m)\big)/2}
\ee
Thus, one of the modes vanishes as $r\rightarrow 0$ provided $\lambda$ is positive and hence
we can get stable attractors upon setting $C_2 = 0$. Unfortunately, the examples we consider
in this paper do not admit a critical point with $\lambda >0$. Thus such fluctuations are not 
stable.

Now we turn to the case of time dependent fluctuations. Since the equation for $\delta\phi^x$
is separable, we try the ansatz
$\de\phi(\hat{r},\hat{t})=f(\hat{r}) e^{i k \hat{t}}$ (with $k$ real) to get the Bessel equation:
\be
\biggl[\hat{r}^2 \dd^2_{\hat{r}}+(m+2) \hat{r} \dd_{\hat{r}}+
(\frac{k^2}{\hat{r}^{2u_0}}-\la)\biggr]f(\hat{r})=0 \ .
\ee
The general solutions for this equation are given by the standard Bessel functions (see, for 
example, \cite{jeffrey2007table}, page 932):
\be
f(X)=\biggl(\frac{X}{2}\biggr)^{\nu_0}\biggl[ C_1 \Ga(1-\nu_\la) J_{-\nu_\la}(X)+ C_2
\Ga(1+\nu_\la) J_{\nu_\la}(X)\biggr],
\ee
where,
\begin{align}\label{xrelatedtor}
X &=\frac{k}{u_0 \hat{r}^{u_0}}, \quad \nu_\la = \frac{\sqrt{(1+m)^2+ 4 \la}}{2 u_0}, \quad \nu_0=
\frac{(1+m)}{2u_0} \ , 
\end{align}
$C_1$ and $C_2$ are arbitrary constants, and
\begin{align}
J_{\nu_\la}(X) &= \biggl(\frac{X}{2}\biggr)^{\nu_\la} \sum_{j=0}^{\infty}\frac{(-1)^j}{j!
\Ga(j+\nu_\la+1)} \biggl(\frac{X}{2}\biggr)^{2j}, \nn\\
J_{-\nu_\la}(X) &=\biggl(\frac{X}{2}\biggr)^{-\nu_\la} \sum_{j=0}^{\infty}\frac{(-1)^j}{j!
\Ga(j-\nu_\la+1)} \biggl(\frac{X}{2}\biggr)^{2j}.
\end{align}
The power series representation is valid in the small $X$ or equivalently,  in the large $r$ regime.
We can rewrite the solution in terms of the Hankel functions
\begin{align}
 J_{\nu_\la}(X) &= \frac{1}{2}(H^1_{\nu_\la}(X)+H^2_{\nu_\la}(X)),\nn\\
 J_{-\nu_\la}(X) &=\frac{1}{2}(H^1_{\nu_\la}(X) e^{i\nu_\la\pi }+H^2_{\nu_\la}(X) e^{-i\nu_\la\pi}
),
\end{align}
to get,
\begin{align}\label{solutionfofx}
f(X)=\biggl(\frac{X}{2}\biggr)^{\nu_0}&\biggl[ C_1
H^1_{\nu_\la}(X)\bigl[\Ga(1-\nu_\la)e^{i\nu_\la\pi }+\Ga(1+\nu_\la) \bigr]\nn\\
&+ C_2 H^2_{\nu_\la}(X)\bigl[\Ga(1-\nu_\la)e^{-i\nu_\la\pi}+\Ga(1+\nu_\la) \bigr]\biggr] \ . 
\end{align}

As one can see from above equation, there is already a restriction on $\nu_\la$ from the Gamma function that
appears in the general solution. First let us consider the case $\nu_\la$ real, then we have the
condition,
\be\label{conditiononlambda1}
\nu_\la=\frac{\sqrt{(1+m)^2+ 4 \la}}{2 u_0}=\frac{\sqrt{(\sum_l c_l u_l)^2+ 4 \la}}{2 u_0} \leq 1,
\ee
for
\be\label{conditiononlambda2}
-\frac{(\sum_l c_l u_l)^2}{4}\leq \la <0.
\ee
Note that only negative $\lambda$ can satisfy \eqref{conditiononlambda1}.\footnote{We do not
consider $\la=0$ as that would leave the nature of the critical point undetermined.}
Since $c_l>0$ and all the
$u_l$ have to be positive for the existence of a regular horizon, we conclude that $\la$ has to be
negative. Remember that the sign of $\la$ is provided by the double derivative of
the attractor potential eqs. (\ref{definitions},\ref{doublederivativeofattractorpotential}). This
implies that the critical points correspond to maxima of the attractor potential. For the case of
imaginary $\nu_\la$ we have,
\be\label{conditionlambda3}
\la < -\frac{(\sum_l c_l u_l)^2}{4},
\ee
and hence, even in this case the critical points correspond to a maxima of the attractor potential.
Thus we have determined the general solution for the scalar fluctuation \eqref{solutionfofx} and
and we find that they are well behaved at large distance provided they satisfy the conditions
 (\ref{conditiononlambda2},\ref{conditionlambda3}). 
This may be useful for
the study of attractor flow equations for black holes in $AdS$.

\subsection{Stable Bianchi attractors}\label{stablebianchiattractors}
In this section we will analyse the stability of the Bianchi attractors by studying the behaviour
of the solution in the $r\rightarrow 0$ limit. We are interested in the
question which class of the Bianchi attractors can be stable attractor geometries in gauged
supergravity. This can be answered by looking at the near horizon behaviour of the scalar
fluctuations \eqref{solutionfofx}. From our analysis of the stress energy tensor in gauged
supergravity \eqref{stressenergytracegeneralgauging} we found that there is dependence on the
fluctuations and their derivatives at first order perturbation. Hence, we only require that the
fluctuations do not blow up near the horizon as that would backreact strongly and deviate from the
geometry. This requirement places some constraints on the form of the metric itself as we explain in
the rest of the section.

Both the solutions in \eqref{solutionfofx} are given in terms of the Hankel functions, the behaviour
near the horizon can be determined by considering the asymptotic expansions of the Hankel functions.
Remember that the horizon for the Bianchi metrics \eqref{bianchimetricgeneralform} is located at
$\hat{r}=0$ The form of the solution \eqref{solutionfofx} makes it convenient to use the asymptotic
expansions of the Hankel functions, since from \eqref{xrelatedtor} $X\rightarrow \infty$ as
$\hat{r}\rightarrow 0$. The asymptotic expansions are given by,
\begin{align}\label{hankelasymptoticexpansion}
 H^1_{\nu_\la}(X) &\sim\sqrt{\frac{2}{\pi X}}e^{i (X-\frac{\pi}{2}(\nu_\la+\frac{1}{2}))}\nn\\
 H^2_{\nu_\la}(X) &\sim\sqrt{\frac{2}{\pi X}}e^{-i (X-\frac{\pi}{2}(\nu_\la+\frac{1}{2}))}.
\end{align}
Substituting \eqref{hankelasymptoticexpansion} in \eqref{solutionfofx} we determine the behaviour of
the fluctuation near the horizon as,
\begin{align}\label{nearhorizonbehavior}
f(X)\sim \biggl(\frac{X}{2}\biggr)^{\nu_0-\frac{1}{2}}\sqrt{\frac{1}{\pi}}&\biggl[ 
C_1 e^{i (X-\frac{\pi}{2}(\nu_\la+\frac{1}{2}))}
\bigl[\Ga(1-\nu_\la)e^{i\nu_\la\pi }+\Ga(1+\nu_\la) \bigr]\nn\\
&+ C_2 e^{-i
(X-\frac{\pi}{2}(\nu_\la+\frac{1}{2}))} \bigl[\Ga(1-\nu_\la)e^{-i\nu_\la\pi}+\Ga(1+\nu_\la) \bigr]
\biggr].
\end{align}
Since $X \sim \frac{1}{\hat{r}^{u_0}}$ and $u_0 > 0$, there is a leading divergent term as $\hat{r}
\rightarrow 0$ unless 
\be
\frac{1-2\nu_0}{2}\geq 0 ,
\ee
which can be rewritten as,
\be
\nu_0=\frac{(1+m)}{2u_0}=\frac{\sum_l c_l u_l}{2 u_0}\leq\frac{1}{2}.
\ee
Since $c_0=1$, this implies,
\be\label{stabilitycondition1}
\sum_{l, l\neq 0} c_l u_l \leq 0,
\ee
which can never be satisfied without some of the exponents $u_l$ being negative. Since we require a
regular horizon, all the exponents have to be positive. Thus the only possibility for which eq. \eqref{stabilitycondition1} can be satisfied is
\be\label{stabilitycondition2}
u_0 \neq 0 ,\quad u_l=0 \ \ \forall \ l \neq 0.
\ee
The conditions on $\la$ \eqref{conditiononlambda2},\eqref{conditionlambda3} for the general solution
\eqref{solutionfofx} to exist can now be written as,
\be
-\frac{u_0^2}{4}\leq \la < 0, 
\ee
for real $\nu_\la$, and
\be
\la<-\frac{u_0^2}{4},
\ee
for imaginary $\nu_\la$. To summarise, Bianchi attractors are stable against scalar fluctuations
about the attractor value for the class of metrics which satisfy the condition \eqref{stabilitycondition2}.

The condition \eqref{stabilitycondition2} is highly restrictive on the form of the Bianchi metrics.
In particular it follows from \eqref{stabilitycondition2} that  $\nu_0=\frac{1}{2}$ for any $u_0>0$
and the scalar fluctuations \eqref{nearhorizonbehavior} do not diverge near the
horizon.\footnote{Note that
there are still oscillatory terms in the fluctuation.} In
particular this
restricts the metrics \eqref{bianchimetricgeneralform} to be of the form, 
\be\label{bianchimetricstableform}
ds^2=L^2 \biggl[-\hat{r}^{2u_0} d\hat{t}^2+ \frac{d\hat{r}^2}{\hat{r}^2}+\eta_{ij}\om^i\otimes
\om^j\biggr].
\ee

It is very interesting to note that the symmetry group of this metric form factorises into a direct
product of the $(1+1)$ dimensional Lifshitz group and a group in the Bianchi classification. This is
similar to what happens for example in four dimensional extremal black holes where the near
horizon geometry factorises as $AdS_2 \times S^2$. 

The simplest non-trivial example of this class is the $Lif_{u_0}(2) \times M_I$ solution, 
\be
ds^2 = L^2\bigg[-\hat{r}^{2 u_0} d{\hat{t}}^2+ \frac{d \hat{r}^2} {\hat{r}^2}+
(d\hat{x}^2+d\hat{y}^2+d\hat{z}^2)\bigg],
\ee 
one obtains the $AdS_2 \times \mathbb{R}^3$ solution when $u_0=1$. 
Another less trivial example is
the $Lif_{u_0}(2) \times M_{II}$ solution,
\be
ds^2 = L^2\bigg[-\hat{r}^{2u_0} d\hat{t}^2+ \frac{d \hat{r}^2} {\hat{r}^2}+
(d\hat{x}^2+ d\hat{y}^2 -2\hat{x}  d\hat{y} d\hat{z}+ (\hat{x}^2+1) d\hat{z}^2) \bigg].
\ee

We have constructed the $Lif_{u_0}(2) \times M_I$ for any $u_0>0$ and a
$Lif_{u_0}(2) \times M_{II}$ in a simple $U(1)_R$ gauged supergravity theory with one
vector multiplet. These solutions and the details of the theory are summarised in Appendix
\eqref{ads2R3ingaugedsugra}. It can be seen from Table \eqref{table1} that these solutions satisfy
the stability criteria \eqref{stabilitycondition2} and hence are examples of stable Bianchi
attractors in gauged supergravity.
\begin{table}
\centering
\begin{tabular}{|c|c|c|c|c| }
\hline
 Geometry & $\la$ & $u_0$ & $u_l ,l\neq 0$ & Stability\\
\hline
 Lifshitz & $-34$ & $3$ & $1$ & no\\ [0.2cm]
\hline
Bianchi II & $-\frac{22}{3}$ & $\sqrt{2}$& $u_1=u_2=\frac{1}{2\sqrt{2}}$ & no\\[0.2cm]
\hline
Bianchi VI $h<0$ & $-1 + \frac{14 h}{3} - h^2$ & $\frac{1}{\sqrt{2}}(1-h)$
&$u_1=-\frac{1}{\sqrt{2}}h, u_2=\frac{1}{\sqrt{2}}$ &no \\[0.2cm]
\hline
$Lif_{u_0}(2) \times M_I $ & $-\frac{5 u_0^2}{3}$ & any $u_0>0$ & $0$ & yes\\[0.2cm]
$AdS_2 \times M_I $  & $-\frac{5}{3}$ & $1$ & $0$& yes\\[0.2cm]
$Lif_{u_0}(2) \times M_{II}$ &$-\frac{61}{6}$& $\sqrt{\frac{11}{2}}$ & $0$ & yes\\[0.2cm]
$Lif_{u_0}(2) \times M$* & $\la<0$ & any $u_0>0$ & $0$& yes\\[0.2cm]
\hline
\end{tabular}
\caption{\em Bianchi attractor geometries in gauged supergravity, nature of critical points and
stability. The first three entries are for the solutions found in \cite{Inbasekar:2012sh}. The next
three entries are generalised attractors in $U(1)_R$ gauged supergravity
\eqref{ads2R3ingaugedsugra}. The last
entry with the * is the most general possible Bianchi attractor geometry
\eqref{bianchimetricstableform} that satisfies our stability criteria.}
\label{table1}
\end{table}

The examples we constructed earlier in \cite{Inbasekar:2012sh} have $\la<0$ and exist at maxima of
the attractor potential. Therefore the condition \eqref{conditiononlambda1} allows scalar
fluctuations about the attractor values. However as one can see from Appendix
\ref{gauged_sugra_solutions} all the metrics have some $u_l\neq 0$ for $l \neq 0$ and do not satisfy
\eqref{stabilitycondition2}. Hence the radial fluctuation of the scalar field diverges near the
horizon for all these metrics. To complicate matters further, as one can see from
\eqref{stressenergytracegeneralgauging} the fluctuations and their derivatives backreact on the
geometry strongly. Thus there would be significant deviation of the geometry even at the 
first order and
we conclude that these geometries are unstable attractors in the theory. These results are
summarised in Table \eqref{table1}.

\section{Summary}\label{summary}
We have studied the stability of Bianchi attractors in gauged supergravity by considering scalar
fluctuations about the attractor value. In general, the stress energy tensor in a generic gauged
supergravity depends on the scalar fluctuations and their derivatives even at first order
perturbation. Therefore, it is important that the scalar fluctuations are well behaved near the
horizon. In particular, if there is a large backreaction then the geometry would deviate from the
attractor geometry. Hence the fluctuations must vanish as one approaches the horizon for the
attractor geometry to be stable.

We analysed the scalar fluctuation equations and found that the fluctuations can exist in general
when the attractor geometries in consideration exist at critical points which, in the present case, 
correspond to maxima of the attractor potential. By demanding that the
fluctuations vanish as one approaches the horizon we determined the conditions of stability for the
metric. We found that the Bianchi attractors are stable if the metric factorises as,
\be
ds^2=L^2 \left(-\hat{r}^{2u_0} d\hat{t}^2+ \frac{d\hat{r}^2}{\hat{r}^2}\right)
+ L^2  \left(\eta_{ij}\om^i\otimes
\om^j\right),
\ee
which is a subclass of the Bianchi attractors constructed by \cite{Iizuka:2012iv}. We refer to
this class of metrics as $Lif_{u_0}(2)\times M$, where $M$ refers to three dimensional manifolds
invariant under the nine groups given by the Bianchi classification. We also constructed explicit
examples of $Lif_{u_0}(2)\times M_I$ and $Lif_{u_0}(2) \times M_{II}$ in a $U(1)_R$
gauged supergravity using the generalised attractor procedure. As stated before, these solutions
exist for critical points which are maxima of the attractor potential and they satisfy all the
conditions of stability. It would be interesting to explore whether this is a generic feature of attractors
in gauged supergravity or an artifact of the models we are considering in this paper.\\{}\\

\noindent
\textbf{\large{Acknowledgements:}} We would like to thank Sandip Trivedi for useful discussions. 
K.I would also like to thank Nemani Suryanarayana, Bala Sathiapalan, Shankhadeep Chakraborty and
Sudipto Paul Chowdhury for useful discussions. The work of K.I is supported by a research
fellowship from the Institute of Mathematical Sciences, Chennai. 

\appendix
\section{Tangent space and constant anholonomy}\label{tangentspaceandanholonomy}

In this section of the appendices we describe notations and conventions used in tangent space. We
use Greek indices  $\mu, \nu,\ldots = 0,1,\ldots, 4$ to denote  denote
the space time coordinates where as Latin indices $a,b,\ldots = 0,1,\ldots, 4$ to denote the tangent
space coordinates.  The tangent space and space time metrics are denoted by $\eta_{ab}$ and 
$g_{\mu\nu}$ respectively with signature $ \{-,+,+,+,+\}$. The
vielbeins $e^a_\mu(x)$ are related to the space time metric $g_{\mu\nu}$ by
\be
g_{\mu\nu}=e^a_\mu e^b_\nu \eta^{ab}.
\ee
The covariant derivative with respect to the spin connection $\omega_{a,bc}$  is denoted 
by $D_a(\omega)$. The covariant derivative is defined in terms of its action on spinors  
$\chi_\alpha$
\be
D_a(\om)\chi_\al= \dd_a\chi_\al-\frac{1}{4}\om_{a}^{\  bc}\ga_{bc}\chi_\al,
\ee
and on vectors $V^a$
\be
D_a(\om) V^b = \dd_a V^b + \omega_{a, \ c}^{\ \ b} V^c.
\ee

We introduce the one forms $e^a\equiv e^a_\mu dx^\mu$ associated with the vielbeins 
$e^a_\mu$ and the corresponding dual vector fields $ \et_a \equiv e^\mu_a \dd_\mu$. The
anholonomy coefficients are defined to be the structure constants of  Lie algebra associated
with the duals vector fields $\et_a$:
\be\label{anholonomy} 
[\et_a,\et_b] \equiv c_{ab}^{\ \ c} \et_c  .
\ee
They can also be expressed in terms of the vielbeins $e^a_\mu$ as 
\be
\quad  c_{ab}^{\ \ c} = e^\mu_a e^\nu_b (\dd_\nu e_\mu^c -\dd_\mu e_\nu^c) \ . 
\ee

In the absence of torsion the spin connection can uniquely be determined in terms of the 
anholonomy coefficients using the relation:
\be\label{anholonomyandspinconnection}
\omega_{a,bc} =\frac{1}{2} [ c_{ab,c}- c_{ac,b}- c_{bc,a}], 
\ee
where $\omega_{a ,bc}=-\omega_{a ,cb} $ and $c_{ab,c}=- c_{ba,c}$. 
It is straightforward to compute the tangent space components of the Riemann tensor. It can be 
written in terms of the anholonomy coefficients and the spin connection as:
\be\label{riemann}
R_{abc}^{\ \ \ d}= \dd_a\omega_{bc}^{\ \ d}-\dd_b\omega_{ac}^{\ \ d}-\omega_{ac}^{\ \
e}\omega_{be}^{\ \ d}+\omega_{bc}^{\ \ e}\omega_{ae}^{\ \ d}-c_{ab}^{\ \ e}\omega_{ec}^{\ \ d}.
\ee
For constant anholonomy coefficients the partial derivatives acting on the spin connections vanish
and as as result the curvature tensor is expressed entirely as a function of the anholonomy 
coefficients.

\section{Gauged supergravity with one vector multiplet:} \label{gauged_sugra_model}

In this section we will discuss one of the simplest gauged supergravity model in five dimensions
coupled to one vector multiplet constructed by Gunaydin and Zagermann
\cite{Gunaydin:1999zx,Gunaydin:2000xk}. 
We will outline some of the important results derived by them which are useful to study the
generalised attractors. This gauged supergravity model consists 
of one gravity multiplet, one vector multiplet and two tensor multiplets with field contents: 
\be 
\{e_\mu^a, \psi_\mu^i, A_\mu^I,B_{\mu\nu}^M, \lambda^{i\at}, \phi^{\xt} \} \ , 
\ee
with the indices taking values such that $i=1,2$ ; $\mu=0,\ldots,4$ ; $a=0,\ldots,4$ ; $I=0,1$ ; 
$M=2,3$ ; $\xt=1,2,3$ and $\at=0,1,2,3$.  Here $\psi_\mu^i$ are the gravitinos, $A_\mu^I$ 
are the vectors in the gravity and vector multiplets, $\lambda^{i\at}$ are the gaugini and 
$B_{\mu\nu}^M$ are antisymmetric tensors in the tensor multiplets. The scalars in the vector 
and tensor multiplet are collectively written as $\phi^{\xt}$. We also use the index $\It$ 
to label the vector and tensor multiplet indices collectively.
 
 In five dimensional supergravity the moduli space of  scalars $h^{\It}\equiv h^{\It}(\phi)$ 
 in the vector and tensor multiplets parametrise a very special manifold $\Sm$ which is 
 described by a cubic surface:
\be\label{scalarmanifold_constraint}
N\equiv C_{\It \Jt \Kt} h^{\It} h^{\Jt} h^{\Kt}=1.
\ee
In the present case the scalar manifold is given by the coset space:
\be
\Sm=SO(1,1)\times \frac{SO(2,1)}{SO(2)}
\ee
The symmetry group of the scalar manifold $\Sm$ is $G= SO(1,1)\times SO(2,1)$.  We can gauge 
a subgroup of this symmetry group $G$ as well as a subgroup of the $R$-symmetry group $SU(2)_R$. One
possibility for gauging is the $SO(2) \subset SO(2,1)$ symmetry of the scalar manifold 
using the graviphoton $A^{0\mu}$ and the subgroup $U(1)_R \subset SU(2)_R$ using $A_{\mu}(U(1)_R)=V_I A^I_\mu$. 

The symmetries of the scalar manifold can be made manifest by going to a suitable basis such that
$h^{\It}=\sqrt{\frac{2}{3}}\xi^{\It}|_{N=1}$ and $h_{\It}=\frac{1}{\sqrt{6}}\frac{\partial}{\partial
\xi^{\It}}N|_{N=1}$. In such a parametrisation, the constraint \eqref{scalarmanifold_constraint}
takes the form
\be
N(\xi)=\sqrt{2} \xi^0 [ (\xi^1)^2-(\xi^2)^2-(\xi^3)^2 ]=1,
\ee
where,
\be
\xi^0= \frac{1}{\sqrt{2} ||\phi||^2} ;\ \ \xi^1=\phi^1 ;\ \ \xi^2=\phi^2 ;\ \xi^3=\phi^3 ,
\ee
and,
\be
||\phi||^2=(\phi^1)^2-(\phi^2)^2-(\phi^3)^2
\ee
is assumed to be positive so that the metrics $a_{\It \Jt}$ and $g_{\xt \yt}$ are positive definite. The
non-vanishing components of $C_{\It \Jt \Kt}$ are given by  $C_{011}=\frac{\sqrt{3}}{2},
C_{022}=C_{033}=-\frac{\sqrt{3}}{2}$. The $h^{\It}$ are related to the fields $\phi$ in the
Lagrangian through the following relations,
\begin{displaymath}
h^{0}=\frac{1}{\sqrt{3}||\phi||^2}, \qquad
h^{1}=\sqrt{\frac{2}{3}}\phi^1,  \qquad
h^{2}=\sqrt{\frac{2}{3}}\phi^2,  \qquad
h^{3}=\sqrt{\frac{2}{3}}\phi^3. \nn
\end{displaymath}
\begin{displaymath}
h_{0}=\frac{1}{\sqrt{3}}||\phi||^2, \qquad
h_{1}=\frac{2}{\sqrt{6}}\frac{\phi^1}{||\phi||^2}, \qquad
h_{2}=-\frac{2}{\sqrt{6}}\frac{\phi^2}{||\phi||^2}, \qquad
h_{3}=-\frac{2}{\sqrt{6}}\frac{\phi^3}{||\phi||^2}.
\end{displaymath}
The moduli space metric has the expression
\be
g_{\xt \yt}= \begin{pmatrix}
 4 (\phi^1)^2 ||\phi||^{-4}-||\phi||^{-2} & -4 \phi^1\phi^2 ||\phi||^{-4} & -4 \phi^1 \phi^3
||\phi||^{-4}\\
-4 \phi^1\phi^2 ||\phi||^{-4} & 4 (\phi^2)^2||\phi||^{-4}+||\phi||^{-2} & 4\phi^2\phi^3
||\phi||^{-4}\\
-4 \phi^1\phi^3 ||\phi||^{-4} & 4\phi^2\phi^3 ||\phi||^{-4} & 4 (\phi^3)^2
||\phi||^{-4}+||\phi||^{-2}
\end{pmatrix}
\ee
The Killing vector that generates the $SO(2)$ symmetry is given by
\be\label{killingvectorSO2}
K_0^{\xt}=\bigg\{ -\frac{\phi^1}{||\phi||^2}, \frac{\phi^2}{||\phi||^2}, \frac{\phi^3}{||\phi||^2}
\bigg\}.
\ee
For convenience, we also give the matrix representation for $a_{\It \Jt}$:
\be
\begin{pmatrix}
 ||\phi||^4 & 0 & 0 & 0 \\
0 &  2 (\phi^1)^2 ||\phi||^{-4}-||\phi||^{-2} & -2 \phi^1\phi^2 ||\phi||^{-4} & -2 \phi^1 \phi^3
||\phi||^{-4}\\
0 & -2 \phi^1\phi^2 ||\phi||^{-4} & 2 (\phi^2)^2||\phi||^{-4}+||\phi||^{-2} & 2\phi^2\phi^3
||\phi||^{-4}\\
0 & -2 \phi^1\phi^3 ||\phi||^{-4} & 2\phi^2\phi^3 ||\phi||^{-4} & 2 (\phi^3)^2
||\phi||^{-4}+||\phi||^{-2}
\end{pmatrix}
\ee
The scalar potential of this model is given by,
\be\label{scalarpotentialinmodel}
\V(\phi)= \frac{g^2}{8} \bigg[ \frac{[(\phi^2)^2+(\phi^3)^2]}{||\phi||^6}\bigg]-2 g_R^2
\bigg[2\sqrt{2} \frac{\phi^1}{||\phi||^2} V_0 V_1 + ||\phi||^2 V_1^2\bigg].
\ee
\subsection{Generalised attractor solutions in Gauged supergravity with one vector
multiplet.}\label{gauged_sugra_solutions}
We summarise the generalised attractor solutions found earlier in \cite{Inbasekar:2012sh} for
reference. Using the generalised attractor procedure, some explicit examples of Bianchi attractors
were constructed within the gauged supergravity model described in the previous section. Firstly,
the vaccum $AdS_5$ solution is given by,
\\{}\\
\fbox{
\begin{minipage}{\linewidth}
\begin{align}\label{AdS5}
 & ds^2 = L^2\bigg[-\hat{r}^{2} d\hat{t}^2+ \frac{d \hat{r}^2} {\hat{r}^2}+
\hat{r}^2(d\hat{x}^2+d\hat{y}^2+d\hat{z}^2)\bigg]\nn\\
 & \phi^2_c=0, \quad \phi^3_c=0,\quad \phi^1_c=\bigg(\sqrt{2}\frac{V_0}{V_1}\bigg)^{\frac{1}{3}},
\quad \Lambda=-6 { g_R^2 V_1^2}({ \phi^1_c})^2,\nn\\
 & V_0 V_1 > 0 ,\quad 32 \frac{g_R^2}{g^2}V_0^2\leq1, \quad L^2 =-\frac{6}{\La},
 \end{align}
\end{minipage}
}\\{}\\
where $\La$ is the cosmological constant. All the Bianchi examples exist at the same critical values
of the scalars for which there is also an $AdS$ solution. All of them are electrical and are sourced
by a single time like gauge field (the
graviphoton $A^{0\mu}$) 
\be\label{gaugefieldansatz}
A^{0\hat{t}}=e^{\hat{t}}_{\bar{0}} A^{0\bar{0}}=\frac{\ \ \hat{r}^{-u_0} }{L} A^{0\bar{0}}.
\ee
The Lifshitz solution is given by \\{}\\
\fbox{
\begin{minipage}{\linewidth}
\begin{align}\label{Lifshitz}
 & ds^2 = L^2\bigg[-\hat{r}^{2u_0} d\hat{t}^2+ \frac{d \hat{r}^2} {\hat{r}^2}+
\hat{r}^2(d\hat{x}^2+d\hat{y}^2+d\hat{z}^2)\bigg],\nn\\
 & u_0=3, \quad L= \sqrt{3} \frac{(\phi^1_c)^4}{g}, \quad
A^{0\bar{0}}=\sqrt{\frac{2}{3}}\frac{1}{(\phi^1_c)^2}, \nn\\
 & \phi^1_c=\bigg(\sqrt{2}\frac{V_0}{V_1}\bigg)^{\frac{1}{3}}, \quad V_0 V_1 > 0 , \quad\frac{32}{3
(\phi^1_c)^4}\leq1.
\end{align}
\end{minipage}
}\\{}\\
The Bianchi II solution is given by,\\{}\\
\fbox{
\begin{minipage}{\linewidth}
\begin{align}\label{solutionII}
 & ds^2 = L^2\bigg[-\hat{r}^{2u_0} d\hat{t}^2+ \frac{d \hat{r}^2} {\hat{r}^2}+
\hat{r}^{2u_2}d\hat{x}^2+
\hat{r}^{2(u_1+u_2)} d\hat{y}^2\nn\\
& \quad\quad\quad\quad\quad\quad-2\hat{x} \hat{r}^{2(u_1+u_2)} d\hat{y} d\hat{z}
+[\hat{r}^{2(u_1+u_2)}\hat{x}^2+\hat{r}^{2u_1}] d\hat{z}^2 \bigg],\nn\\
  &u_0=\sqrt{2},\quad u_1=u_2=\frac{1}{2\sqrt{2}},\quad L= \sqrt{\frac{2}{3}}
\frac{(\phi^1_c)^4}{g},
\quad A^{0\bar{0}}=\sqrt{\frac{5}{8}}\frac{1}{(\phi^1_c)^2}, \nn\\
  &\phi^1_c=\bigg(\sqrt{2}\frac{V_0}{V_1}\bigg)^{\frac{1}{3}}, \quad V_0 V_1 > 0 , \quad\frac{23}{2
(\phi^1_c)^4}\leq1.
\end{align}
\end{minipage}
}\\{}\\
whereas the Bianchi VI solution is given by,\\{}\\
\fbox{
\begin{minipage}{\linewidth}
\begin{align}\label{solutionVI}
 & ds^2 = L^2\bigg[-\hat{r}^{2u_0} d\hat{t}^2+ \frac{d \hat{r}^2} {\hat{r}^2}+ d\hat{x}^2 +
e^{-2\hat{x}} \hat{r}^{2u_1} d\hat{y}^2+e^{-2h\hat{x}} \hat{r}^{2u_2} d\hat{z}^2\bigg],\nn\\
& u_0=\frac{1}{\sqrt{2}}(1-h),\quad u_1=-\frac{1}{\sqrt{2}}h,\quad u_2=\frac{1}{\sqrt{2}},\quad L=
\frac{(\phi_1^c)^4}{\sqrt{6}g}(1-h),\quad \nn\\
& A^{0\bar{0}}=\sqrt{\frac{-2h}{(-1+h)^2}}\frac{1}{(\phi^1_c)^2},\quad h<0, \quad h\neq0,1, \nn\\
&\phi^1_c=\bigg(\sqrt{2}\frac{V_0}{V_1}\bigg)^{\frac{1}{3}}, \quad V_0 V_1 > 0, \quad
\frac{8(3-h+3h^2)}{(\phi^1_c)^4(-1+h)^2}\leq 1
\end{align}
\end{minipage}
}\\{}\\
\section{Bianchi attractors in \texorpdfstring{$U(1)_R$} a gauged
supergravity}\label{ads2R3ingaugedsugra}
We consider a truncated version of the gauged supergravity model discussed earlier
with just $U(1)_R$ gauging. There is no gauging of the symmetries of the scalar manifold and hence
there are no tensors as well. The field content of the reduced model is given by,
\be 
\{e_\mu^a, \psi_\mu^i, A_\mu^I, \lambda^{i\at}, \phi^1 \},
\ee
with $I=0,1$ and $I=0$ corresponds to the graviphoton as before. The field $\phi^1$ is the scalar
in the single vector multiplet. The gauge field combination used for the $U(1)_R$ gauging is same as
before. The potential of the $U(1)_R$ gauged supergravity contains only terms proportional to
$g_R^2$ as there is no gauging of the symmetries of scalar manifold and can be obtained by setting
$\phi^2=\phi^3=0$ in \eqref{scalarpotentialinmodel},
\be
\V(\phi^1)= -2 g_R^2\bigg[ \frac{2\sqrt{2}V_0 V_1}{\phi^1}  + (\phi^1)^2 V_1^2\bigg].
\ee
We show the embedding of the $Lif_{u_0}(2) \times M_I$ solution, which has the form,
\be\label{Lif2BianchiI}
ds^2 = L^2\bigg[-\hat{r}^{2 u_0} d{\hat{t}}^2+ \frac{d \hat{r}^2} {\hat{r}^2}+
d\hat{x}^2+d\hat{y}^2+d\hat{z}^2\bigg]. 
\ee
When $u_0=1$ we have the familiar $AdS_2 \times \mathbb{R}^3$ solution. We consider the gauge field
ansatz as before,
\be\label{gaugefieldansatztwofields}
A^{I\hat{t}}=e^{\hat{t}}_{\bar{0}} A^{I \bar{0}}= \frac{1}{L \hat{r}}A^{I \bar{0}}.
\ee
For the $U(1)_R$ gauged supergravity the field equations at the attractor point for the gauge field,
scalar field and Einstein equation are read off by setting $g=0$ in the corresponding field
equations found for the general gauging considered in \cite{Inbasekar:2012sh}. The gauge field
equation has the form,
\be
a_{I J}[\omega_{a,\ c}^{\ \ a} F^{cb J}+\omega_{a,\ c}^{\ \ b} F^{ac J}]=0,
\ee
and is identically satisfied for the gauge field ansatz considered above.
The scalar field equation is given by,
\be\label{scalarattractor}
\frac{\dd}{\dd\phi^1}\bigg[\V_{attr}(\phi^1,A^{1\bar{0}},A^{0\bar{0}}) \bigg]=0, \quad
\V_{attr}(\phi^1,A^{1\bar{0}},A^{0\bar{0}})=\V(\phi^1)+\frac{1}{4} a_{IJ} F^I_{ab}F^{J ab}.
\ee
At the critical point $\phi^1_c=(\sqrt{2}\frac{V_0}{V_1})^{\frac{1}{3}}$, it relates the parameters
$V_0$ and $V_1$ to the charges,
\be\label{chargerelation}
(A^{1\bar{0}})^2 - 2 (A^{0\bar{0}})^2 (\phi^1_c)^6=0.
\ee
The Einstein's equations are,
\be
R_{ab}-\frac{1}{2}R \eta_{ab}= T_{ab}^{attr},
\ee
where
\be
T_{ab}^{attr}=\V_{attr}(\phi^1,A^{1\bar{0}},A^{0\bar{0}})\eta_{ab}-a_{IJ}
F_{ac}^I F_{b}^{\ c J}.
\ee
At the attractor point, there are only two independent equations in the above set,
\begin{align}
3 (A^{0 \bar{0}})^2 u_0^2 (\phi^1_c)^5 - 12\sqrt{2} L^2 g_R^2 V_0 V_1 &=0 ,\nn\\
u_0^2 \phi^1_c (-2 + 3 (A^{0 \bar{0}})^2 (\phi^1_c)^4) + 12 \sqrt{2} L^2 g_R^2 V_0 V_1 &=0,
\end{align}
Where we have used \eqref{chargerelation} for simplification. This can be solved to get,
\begin{align}
& L^2= - \frac{u_0^2}{2 \Lambda},\quad \Lambda=-6 g_R^2 V_1^2 (\phi^1_c)^2,\nn\\
& A^{0 \bar{0}} =\frac{1}{\sqrt{3} (\phi^1_c)^2},\quad A^{1 \bar{0}}=\sqrt{\frac{2}{3}} \phi^1_c,
\end{align}
where $\Lambda$ is the $AdS$ cosmological constant. Note that the Einstein equation does not place
additional constraints on any of the gauged
supergravity parameters $V_0, V_1, g_R$, unlike in the other Bianchi cases considered here. We
summarise the solution,
\\{}\\
\fbox{
\begin{minipage}{\linewidth}
\begin{align}
& ds^2 = L^2\bigg[-\hat{r}^{2 u_0} d{\hat{t}}^2+ \frac{d \hat{r}^2} {\hat{r}^2}+
d\hat{x}^2+d\hat{y}^2+d\hat{z}^2\bigg] ,\nn\\
& A^{0 \hat{t}}=\frac{1}{L \hat{r}} A^{0 \bar{0}}, \quad  A^{1 \hat{t}}=\frac{1}{L \hat{r}} A^{1
\bar{0}}, \quad
\frac{A^{0 \bar{0}}}{A^{1 \bar{0}}}=\frac{1}{2}\frac{ V_1}{ V_0}
, \quad L^2= - \frac{u_0^2}{2 \Lambda}, \nn\\
&\Lambda=-6 { g_R^2 V_1^2}({ \phi^1_c})^2 ,\quad {
\phi^1_c}=\bigg(\sqrt{2}\frac{ V_0}{ V_1}\bigg)^{\frac{1}{3}} ,\quad V_0 V_1 > 0.
\end{align}
\end{minipage}
}\\{}\\
Note that the solution exists for any $u_0>0$ and in particular, when $u_0=1$ we get the familiar
$AdS_2 \times \mathbb{R}^3$ solution.

The $Lif_{u_0}(2) \times M_{II}$ metric has the form,
\begin{align}\label{Lif2BianchiII}
 & ds^2 = L^2\bigg[-\hat{r}^{2u_0} d\hat{t}^2+ \frac{d \hat{r}^2} {\hat{r}^2}+
d\hat{x}^2+ d\hat{y}^2 -2\hat{x}  d\hat{y} d\hat{z}+ (\hat{x}^2+1) d\hat{z}^2 \bigg].
\end{align}
We consider the same gauge field ansatz as for the previous case \eqref{gaugefieldansatztwofields}.
As earlier, the gauge field equations are identically satisfied. The attractor equations are again
same as \eqref{chargerelation} at the critical point. There are three independent Einstein
equations given by,
\begin{align}
 \phi^1_c +6 (A^{0 \bar{0}})^2 (\phi^1_c)^5-24\sqrt{2} L^2 g_R^2 V_0 V_1 &=0, \nn\\
 \phi^1_c - 4 u_0^2 \phi^1_c+ 4 (A^{0 \bar{0}})^2 u_0^2 (\phi^1_c)^5+ 24\sqrt{2} L^2 g_R^2 V_0 V_1
&=0, \nn\\
-3\phi^1_c-4 u_0^2 \phi^1_c+6 (A^{0 \bar{0}})^2 u_0^2 (\phi^1_c)^5 +24\sqrt{2} L^2 g_R^2 V_0 V_1
&=0,
\end{align}
where we have again used \eqref{chargerelation} for simplification. This set of algebraic equations
can be solved to get,
\begin{align}
&A^{0 \bar{0}}= \frac{\sqrt{2}}{u_0 (\phi^1_c)^2}, \quad A^{1 \bar{0}}= \frac{2 \phi^1_c}{u_0},
\nn\\
& u_0=\sqrt{\frac{11}{2}},\quad L^2 =-\frac{13}{4 \La},
\end{align}
where $\La$ is the AdS cosmological constant. We summarise the $Lif_{u_0}(2) \times M_{II}$ as,
\\{}\\
\fbox{
\begin{minipage}{\linewidth}
\begin{align}
& ds^2 = L^2\bigg[-\hat{r}^{2u_0} d\hat{t}^2+ \frac{d \hat{r}^2} {\hat{r}^2}+
d\hat{x}^2+ d\hat{y}^2 -2\hat{x}  d\hat{y} d\hat{z}+ (\hat{x}^2+1) d\hat{z}^2 \bigg] ,\nn\\
& A^{0 \hat{t}}=\frac{1}{L \hat{r}} A^{0 \bar{0}}, \quad  A^{1 \hat{t}}=\frac{1}{L \hat{r}} A^{1
\bar{0}}, \quad
\frac{A^{0 \bar{0}}}{A^{1 \bar{0}}}=\frac{1}{2}\frac{ V_1}{ V_0}
, \quad u_0=\sqrt{\frac{11}{2}} , \nn\\
&L^2= -\frac{13}{4 \La},\quad \Lambda=-6 { g_R^2 V_1^2}({ \phi^1_c})^2 ,\quad {
\phi^1_c}=\bigg(\sqrt{2}\frac{ V_0}{ V_1}\bigg)^{\frac{1}{3}} ,\quad V_0 V_1 > 0.
\end{align}
\end{minipage}
}\\{}\\
\providecommand{\href}[2]{#2}\begingroup\raggedright\endgroup
\end{document}